\documentclass[letter]{aa} 
\usepackage[varg]{txfonts}
\usepackage{graphicx} 
\usepackage{natbib} 
\bibpunct{(}{)}{;}{a}{}{,} 
\begin{document}

\title{Carbon monoxide in the solar atmosphere} \subtitle{II.
  Radiative cooling by CO lines}
\titlerunning{Carbon monoxide in the solar atmosphere}

\author{S.~Wedemeyer-B\"ohm \inst{1,3} \and M.~Steffen \inst{2} }
 
\offprints{msteffen@aip.de}
 
\institute{Kiepenheuer-Institut f\"{u}r Sonnenphysik,
  Sch\"{o}neckstra\ss e~6, 79104~Freiburg, Germany 
  \and
  Astrophysikalisches Institut Potsdam, An der Sternwarte~16,
  14482~Potsdam, Germany 
  \and 
  Institute of Theoretical Astrophysics, University of Oslo,
  P.O. Box 1029 Blindern, N-0315 Oslo, Norway
} \date{Received 3 August 2006 / Accepted 6 December 2006}

\abstract{}{
  The role of carbon monoxide as a cooling agent for the thermal
  structure of the mid-photospheric to low-chromospheric layers of the
  solar atmosphere in internetwork regions is investigated.  }
{
  The treatment of radiative cooling via spectral lines of carbon
  monoxide (CO) has been added to the radiation chemo-hydrodynamics
  code CO5BOLD.  The radiation transport has now been solved in a
  continuum band with Rosseland mean opacity and an additional band
  with CO opacity. The latter is calculated as a Planck mean over the
  CO band between 4.4 and 6.2\,$\mu$m.  The time-dependent CO number
  density is derived from the solution of a chemical reaction network.
} {
  The CO opacity indeed causes additional cooling at the fronts of
  propagating shock waves in the chromosphere.  There, the
  time-dependent approach results in a higher CO number density
  compared to the equilibrium case and hence in a larger net radiative
  cooling rate.  The average gas temperature stratification of the
  model atmosphere, however, is only reduced by roughly 100\,K.  Also
  the temperature fluctuations and the CO number density are only
  affected to small extent.  A numerical experiment without dynamics
  shows that the CO cooling process works in principle and drives the
  atmosphere to a cool radiative equilibrium state.  At chromospheric
  heights, the radiative relaxation of the atmosphere to a cool state
  takes several 1000 s. The CO cooling process thus would seem to be
  too slow compared to atmospheric dynamics to be responsible for
  the very cool temperature regions observed in the solar atmosphere.
} {
  The hydrodynamical timescales in our solar atmosphere model are much
  too short to allow for the radiative relaxation to a cool state,
  thus suppressing the potential thermal instability due to carbon
  monoxide as a cooling agent. Apparently, the thermal structure and
  dynamics of the outer model atmosphere are instead determined
  primarily by shock waves.  }

 
\maketitle

\section{Introduction}
\label{sec:intro}

Carbon monoxide (CO) has been observed in the solar atmosphere
\citep[e.g.,][]{noyes72b, ayres81, ayres86, solanki94, uitenbroek94}
for more than 30 years now. The discovery of very low temperatures in
extreme limb observations of strong CO lines at 5\,$\mu$m by Noyes \&
Hall provoked a crisis in our understanding of the thermal structure
of the outer layers of the Sun and solar-type stars
\citep{johnson73, wiedemann94},
since the very existence of CO molecules in significant amounts
requires much lower gas temperatures than predicted by standard
semi-empirical models
\citep{val81, fal93}.
Due to its infrared emission, carbon monoxide is known to be a strong
cooling agent that can actively affect the thermal state of the
atmospheric gas itself. Therefore, the idea that CO was capable of
inducing a thermal bifurcation of the solar atmosphere has attracted
considerable attention during the past three decades
\citep[cf.][]{ayres81, kneer83, muchmore85, muchmore86, steffen88, anderson89c}.

Carbon monoxide molecules can form in significant amounts under the
conditions of the solar photosphere and chromosphere. At a given gas
density, the equilibrium CO concentration is a function of
temperature, such that the molecules are quickly dissociated towards
higher temperatures. Spectral lines of CO at infrared wavelengths can
emit radiation effectively and thus cool the gas, causing a cooling
instability.  In the absence of compensating mechanical heating, a
small perturbation towards lower temperature would already allow the
formation of more CO molecules whose infrared lines would further cool
the gas and thus induce the formation of even more CO, ultimately
leading to a "cooling catastrophe"
\citep[cf.][]{ayres81}. 
This way CO can force a cool equilibrium temperature in regions where
radiative equilibrium (RE) conditions prevail, whereas a critical
amount of mechanical heating can cause a transition to a hot state
\citep[e.g.,][hereafter AA89]{anderson89b}. 

In the case of the Sun, the consequences of the CO cooling instability
have been investigated by a simplified time-dependent approach
\citep{steffen88, muchmore85, muchmore88}
that treated CO in instantaneous chemical equilibrium (ICE).  
The ICE assumption was also made by 
\citet{uitenbroek00a}
for computing the spatial distribution of CO in a two-dimensional
slice from a 3D radiation hydrodynamics simulation.  However, this
assumption is only valid as long as the timescale governing the
formation of carbon monoxide is short compared to the relevant
dynamical time scales.  A time-dependent treatment was presented by
\citet[][]{asensio03}, 
but their simulation was restricted to one spatial dimension in order
to keep the problem computationally tractable.  

Recently, detailed 2D simulations of the distribution of carbon
monoxide in the solar atmosphere
\citep[][hereafter Paper~I]{wedemeyer05a}
were carried out with the radiation hydrodynamics code 
\mbox{CO$^5$BOLD}
\citep{freytag02, wedemeyer04a},  
which can treat chemical reaction networks time-dependently. It was
shown that the bulk of CO is located in cooler regions in the middle
photosphere and that a large fraction of all carbon atoms is bound in
carbon monoxide in the layers above, with the exception of hot
propagating shock waves (see Fig.~\ref{fig.xz}). In contrast to the
ICE approach, the high shock temperatures do not cause instantaneous
destruction of CO but rather gradual dissociation of the molecules on
chemical timescales. Clearly, the highly dynamic and intermittent
nature of the solar chromosphere makes the time-dependent approach
mandatory for studying the influence of CO cooling.

The new two-dimensional simulations presented here were done with an
upgraded code version that now accounts for radiative cooling by CO
lines, using the simplified treatment by 
\citet[][hereafter SM88]{steffen88}
and
\citet[][hereafter MU85]{muchmore85}. 
We describe the method and the new simulations in
Sects.~\ref{sec:method} and \ref{sec:sim}, respectively, and present
the results in Sect.~\ref{sec:result}, followed by discussion and
conclusions in Sect.~\ref{sec:discus}.

\section{Method} 
\label{sec:method}

The applied radiation chemo-hydrodynamics code is described in more
detail in Paper~I.  The only difference to the simulations presented
in Paper~I concerns the radiative transfer. Instead of the grey, i.e.,
frequency-independent, approach we use two frequency bands as described
by SM88.  The first band uses the grey Rosseland opacity
$\kappa_\mathrm{R}$ constructed from OPAL/PHOENIX data \citep{opal,
  hauschildt97}.  It excludes the wavelength region around the CO
fundamental vibration-rotation band in the infra-red beyond  
4.6~$\mu\mathrm{m}$ that is accounted for in the second band. The
opacity in the second band consists of the grey Rosseland opacity
$\kappa_\mathrm{R}$ and an additional 
Planck mean CO opacity $\kappa_\mathrm{CO}$ as a function of 
gas temperature and CO number density.  The latter results from
the preceding solution of the chemical reaction network (see Paper~I).

\section{Simulation} 
\label{sec:sim}

The same initial model as in Paper~I is used so that the simulations
with and without CO cooling can be compared directly. The model
consists of 120 horizontal by 140 vertical grid cells with a total
extent of 4800\,km by 2500\,km. It describes a small portion of the
solar surface layers, just large enough to fit in a few granulation
cells. The chemical composition is the same for each cell in the
initial model. The simulation was advanced for 86000\,s of solar time.
The results presented here only refer to the last 50000\,s, whereas
the first 36\,000 s are reserved as a relaxation phase to ensure the
decay of possible initial perturbations.  The lower and upper
boundaries are located at heights $z = -1479$\,km and $z = 1021$\,km,
respectively, where the origin $z = 0$\,km is defined by the average
Rosseland optical depth of unity.  A corresponding three-dimensional
simulation is currently in production 
(\citeauthor{wedemeyer05b}, in press).

\section{Results} 
\label{sec:result}

\begin{figure}[tp] 
\centering 
  \resizebox{\hsize}{!}{\includegraphics{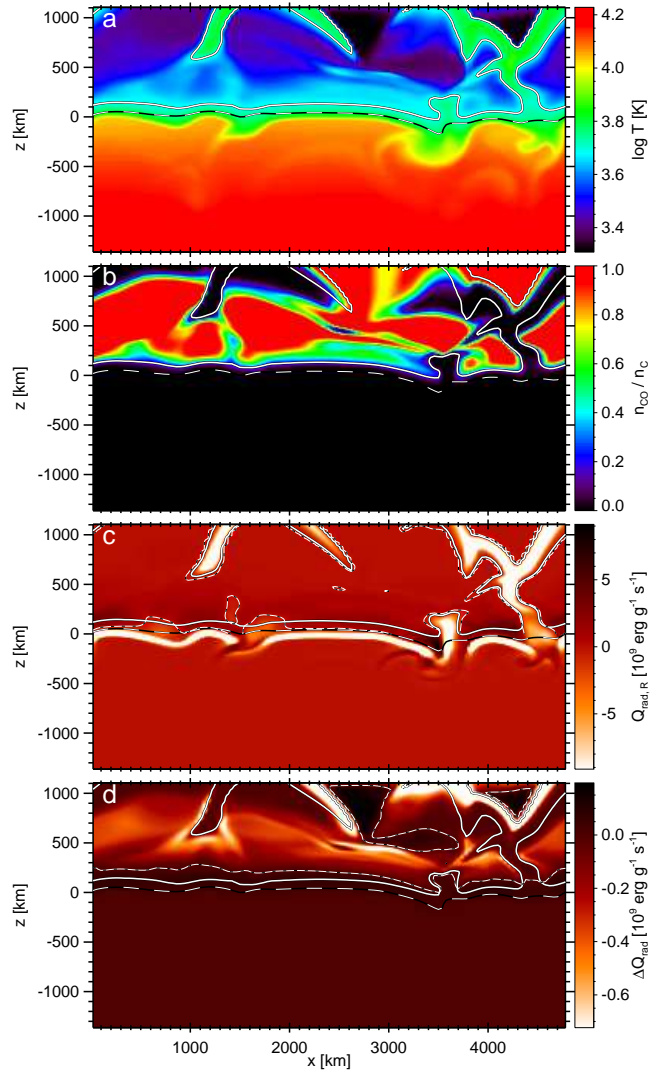}}
  \caption{Two-dimensional slices for an exemplary time step at
    \mbox{$t = 70390$~s}: \textbf{a)} logarithmic gas temperature
    $\log T$, \textbf{b)} fraction of carbon atoms bound in CO
    $n_\mathrm{CO} / n_\mathrm{C, total}$, \textbf{c)} net radiative
    heating rate $Q_\mathrm{rad, R}$ for the simulation without CO
    line cooling, and \textbf{d)} the relative difference $\Delta
    Q_\mathrm{rad}$ between the runs with and without CO cooling.
    Note that the colour range has been limited in the last panel in
    order to to make smaller variations visible. The full range is
    $\Delta Q_\mathrm{rad} = [-2.54,
    0.20]\,10^9\,\mathrm{erg\,g}^{-1}\,\mathrm{s}^{-1}$.  The lines
    represent contours for $T = 5000$~K (white solid) and Rosseland
    optical depth unity (long-dashed).  The short-dashed line in
    panels c and d marks $Q_\mathrm{rad, R} = 0$ and $\Delta
    Q_\mathrm{rad} = 0$, respectively.  }
  \label{fig.xz} 
\end{figure} 
\begin{figure}[t] 
\centering 
  \resizebox{\hsize}{!}{\includegraphics{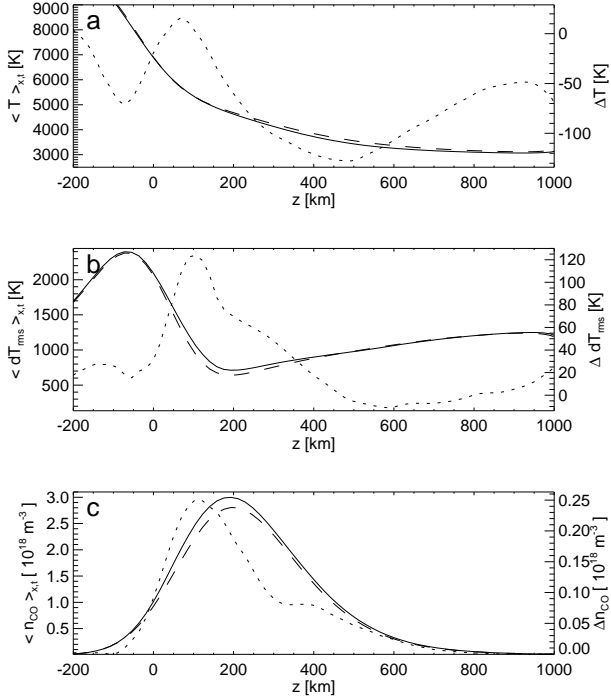}}
  \caption{Horizontally and temporally averaged temperature stratification (\textbf{a}),  
           temperature fluctuation (\textbf{b}), and CO number density (\textbf{c}) 
           for the simulations with (solid line) and 
           without (dashed line) radiative CO cooling. The dotted lines represent the 
           difference between both cases, e.g. 
           $\Delta T = <T_\mathrm{CO}>_{x,t} - <T_\mathrm{R}>_{x,t}$.} 
  \label{fig.strat} 
\end{figure} 

\subsection{Net radiative heating rate}

Direct comparison of the two-dimensional simulations with (this paper)
and without (Paper~I) time-dependent CO opacity shows that inclusion
of the spectral band of CO causes additional cooling at the fronts of
propagating shock waves in the chromosphere. There, the time-dependent
approach results in a higher CO number density compared to the
equilibrium case and thus in a higher net radiative cooling rate.
These small differences make the simulations drift apart so that 
only a statistical comparison makes sense. However, we
illustrate the effect of CO line cooling by comparing the radiative
heating rates calculated with and without additional CO line opacity
for a given model structure (see Fig.~\ref{fig.xz}).  The chosen
snapshot is analysed in more detail in Paper~I.

For reference, panels a and b of the figure show gas temperature and
the fraction of carbon atoms bound in CO, $n_\mathrm{CO} /
n_\mathrm{C, total}$, respectively. The latter refers to the total number
density of all carbon atoms, which is here the sum of the densities of
included species C, CO, and CH. The solid line is a contour for a
temperature of $T = 5000$\,K that outlines the prominent shock wave in
the low chromosphere and also a high-temperature structure belonging
to the reversed granulation pattern in the middle photosphere.

The net radiative heating rate $Q_\mathrm{rad, R}$ for the simulation
without radiative CO cooling is displayed in panel~c. The rate
represents the change in the internal energy of the gas per unit time
and mass due to radiation. Positive values indicate that energy is
added, i.e., the gas is heated, whereas negative values indicate the
release of energy via radiative emission, i.e. radiative cooling. For
most of the computational domain, $Q_\mathrm{rad, R}$ is close to zero
or at least small. In contrast, high negative values are found in
shock waves and close to optical depth unity (the visible ``surface''
of the Sun) where strong radiative emission cools the gas. There are
also some locations with high positive values, indicating radiative
heating of the gas. This is obviously the case for volume elements
that are surrounded by high-temperature regions, e.g. at $x =
3500$\,km, $z = -100$\,km.

Panel~d of Fig.~\ref{fig.xz} shows the additional contribution of CO
to the net radiative heating rate, $\Delta Q_\mathrm{rad}$, which was
retrieved by subtracting the rates from the radiative transfer
calculation with and without additional CO opacity ($\Delta
Q_\mathrm{rad} = Q_\mathrm{rad, CO} - Q_\mathrm{rad, R}$).  Most
obvious is the emission (white) at the fronts of the shock waves where
a significant amount of CO can still effectively cool before being
dissociated by the approaching hot wave fronts. The effect is much
smaller outside shock waves. On the other hand, there are cool regions
with a high CO concentration that absorbs radiation, resulting in a
locally enhanced heating of the gas.  Although this can make a local
difference in the overall contribution of the CO lines, which are
restricted to a relatively narrow wavelength range in the infrared,
the overall effect remains minor compared to the continuum
contribution.

\subsection{Average stratification}

The comparison of the full sequence duration is done by means of
average stratifications as shown in Fig.~\ref{fig.strat} for gas
temperature, root mean square (rms) temperature fluctuation, and CO
number density. The average temperature of the simulation with CO
radiative cooling is a bit lower with a maximum deviation of -128\,K
at a height of 495\,km, whereas the average temperature fluctuation is
increased by maximum 124\,K at $z = 105$\,km. The higher fluctuations
cause a small increase of $0.25\,\cdot\,10^{12}\,\mathrm{cm}^{-3}$ 
(11\,\%) in average CO number density also at a height of 105\,km.

\begin{figure}[t] 
\centering 
  \resizebox{\hsize}{!}{\includegraphics{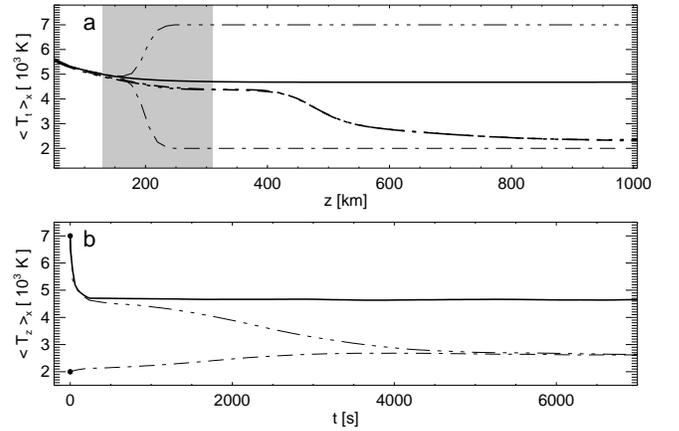}}
  \caption{Time evolution towards radiative equilibrium (RE) for
    simulations without (solid) and with CO line cooling for different
    initial chromospheric temperatures of 2000\,K (dot-dashed) and
    7000\,K (triple-dot-dashed): (\textbf{a}) horizontally averaged
    temperature stratification for initial time step (thin) and after
    10\,000\,s (thick) and (\textbf{b}) time evolution of the
    horizontally averaged temperature at $z = 650$\,km.  The shaded
    area marks the transition layer for artificial damping of velocity
    fluctuations.  }
  \label{fig.coolexp} 
\end{figure} 

\subsection{Cooling experiment} 
\label{sec:coolexp}

The surprisingly small effect of the CO cooling on the thermal
structure of the atmosphere can be understood from the results of a
numerical experiment. As the strong dynamics in the atmosphere disturb
the CO formation, we run additional simulations in which we damp
velocity fluctuations above the middle photosphere by a ``drag-force''
characterised by a timescale of $\tau_\mathrm{damp} = 3$\,s.  After
each time step $\delta t$, all spatial velocity components $v_j$ are
reduced with a reduction factor $r$ that ensures a smooth transition
between the lower undamped domain ($z \le z_1=130$~km) and the upper
damped layers ($z \ge z_2=310$~km):
\begin{eqnarray}
 v_j^\mathrm{new} (i_x, i_z)\ &=&\  r\, v_i^\mathrm{old} (i_x, i_z),
   \enspace \forall j\\
 r\ &=&\  1 - \frac{\delta t\, \zeta^2\, (3\, -\, 2 \zeta)}{\tau_\mathrm{damp}},
   \enspace r \ge 0\\
 \zeta\ &=&\ \frac{z\,(i_z) - z_1}{z_2-z_1},
   \enspace 0\,\ge\,\zeta\,\ge\,1
   \enspace.
\end{eqnarray}
This way the convection in the lower part ($\zeta=0$) remains
unaffected ($r=1$), while the chromosphere ($\zeta=1$) can relax to
hydrostatic, chemical, and radiative equilibrium.  The initial models
are constructed from the reference model without additional CO cooling
(see Paper~I).  The lower part of the model ($z < z_1$) is unchanged.
Above we smoothly attach a new atmosphere with the gas density set to
the horizontal average of the density in the original model and the
gas temperature changed to a constant value of $2000$\,K or $7000$\,K
(see Fig.~\ref{fig.coolexp}.a), which are roughly the extreme values
encountered in the model chromosphere.  The initial abundances of the
molecular species including CO are again set to very small numbers.

Now we run the first simulation with only grey radiative transfer with
the mean Rosseland opacity band, i.e. frequency-independent without CO
band, to derive a reference model in grey radiative equilibrium.  The
resulting temperature stratification (see black line in
Fig.~\ref{fig.coolexp}a) reaches an equilibrium value of 4680\,K in
the upper photosphere and above. The next simulations with the
additional CO band produce the same temperature stratification up to
$z \sim 150$\,km.  Above that height, CO line cooling reduces the mean
gas temperature significantly. For instance, an equilibrium value of
$\sim 3300$\,K is reached at a height of 500\,km, whereas at 650\,km
the final value is close to 2650\,K, independent of the initial
chromospheric temperature.

The temporal evolution of the average temperature is shown for the
chromospheric height of 650\,km in panel~b. Even for high initial
temperature, the grey radiative equilibrium is reached after only a
few minutes.  Also the simulation with the cool initial chromosphere
(2000\,K) shows a steeper evolution towards the grey RE at the very
beginning, but is dominated by the relaxation towards the CO-driven
cool state afterwards.  This second part is much slower and
approximately exponential with timescales of at least $\sim 1000$\,s.
The total time for thermal relaxation to a cool state is of the order
of 3000\,s.  The hot initial chromosphere evolves even slower towards
the equilibrium state.  In all cases the relaxation timescales are
long compared to the hydrodynamical timescales present in the
simulations.

\subsection{Prescribed mechanical heating}
\label{sec:presheat}

\begin{figure}[t] 
\centering 
  \resizebox{\hsize}{!}{\includegraphics{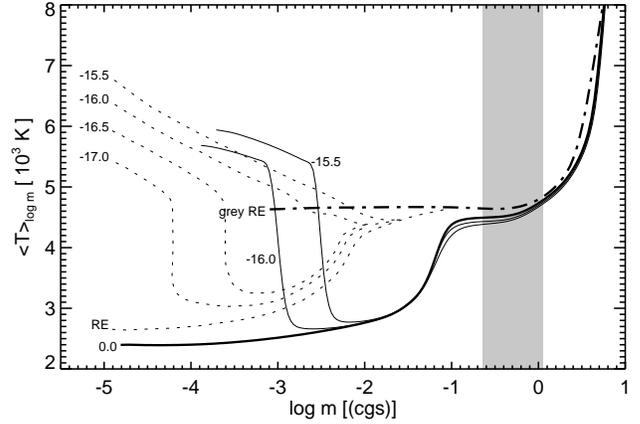}}
  \caption{
    Cooling experiment with prescribed (mechanical) heating.  Gas
    temperature after 4\,h averaged on column mass scale for different
    amounts of prescribed (mechanical) heating with $\log q_0 = -15.5$
    and $-16.0$ (thin solid) and without additional heating (thick
    solid).  The thick dot-dashed line represents a simulation with
    grey radiative transfer alone (i.e. without CO).  The results of
    AA89 (see their Fig.~1) are plotted for comparison (dotted lines).
    The corresponding mechanical heating ($\log q_0$) is noted on the
    left.  The curve annotated with ``RE'' is the non-LTE RE
    stratification by \citet{anderson89c}.  The shaded area marks the
    transition layer for artificial damping of velocity fluctuations.
  }
  \label{fig.presheat} 
\end{figure} 

We extended the cooling experiment (see Sect.~\ref{sec:coolexp}) and
added a prescribed (mechanical) heating as done by AA89 (cf. their
Eq.~(1)).  The additional heating $Q_\mathrm{add}$ [erg/g/s] is a
function of column mass density $m$ :
\begin{equation}
  Q_\mathrm{add}\,(m)\ =\ \mu^{-1}\,q_0\, (m / m_1)^x\, e^{-m/m_0}\enspace, 
\end{equation}
where $q_0$ prescribes the heating rate, $m_0 = m_1 = 0.0114$ and $x =
-0.75$.  The mean atomic weight $\mu$ is set to $2.167\cdot
10^{-24}\,\mathrm{g/atom}$.  The heating term is added to the internal
energy $\varepsilon$ for each grid cell for each time step of duration
$\delta t$:
\begin{equation}
  \varepsilon^\mathrm{new}\,(i_x, i_z)\ =\ 
  \varepsilon^\mathrm{old}\,(i_x, i_z) + 
  \delta t\, Q_\mathrm{add}\,(i_x, i_z)\enspace.
\end{equation}
The initial model is the same as for the simulation described in
Sect.~\ref{sec:sim}.  The calculations are carried out until the
temperature at all heights reaches equilibrium.  The resulting
temperature stratifications are shown in Fig.~\ref{fig.presheat} for
different values of $q_0$.  For comparison the non-LTE models by AA89
are plotted.  We find the same qualitative dependence of equilibrium
temperature stratification on the amount of prescribed heating.  Our
models have a ``temperature minimum'' in the upper photosphere/low
chromosphere and a steep temperature jump above. Nevertheless there
are quantitative differences: (i) Our minimum temperatures and also
the stratifications for no additional heating are cooler than those of
AA89, (ii) the temperature jump occurs deeper in the atmosphere than
for AA89, and (iii) the amount of heating required for compensating
the CO cooling is higher compared to AA89.  In their calculations the
otherwise present cool atmospheric layer cannot form once the heating
exceeds a critical value that lies between $\log q_0 = -16.0$ and
$-15.5$ (see their Fig.~1).  In contrast, we still get a temperature
minimum with $\log q_0 = -15.5$, suggesting that the CO cooling to be
compensated is higher in our models.  The different height dependence
of the average temperature profile is most likely due to differences
in the treatment of radiative transfer.

\section{Discussion and conclusions}
\label{sec:discus}

Our cooling experiment produces a total time for thermal relaxation to
a cool state of the order of 3000\,s at low-chromospheric heights. This
result is in line with the earlier results from SM88.  The final
temperature, however, remains $\sim 350$\,K higher in our experiment.
This can be partly explained by the use of a detailed chemical
reaction network instead of equilibrium CO densities and the
corresponding uncertainties in chemical input data in both cases.
Remaining differences can be attributed to the fact that the
simulations by Steffen \& Muchmore include a stationary flow that
differs from the flow field in our simulations and the resulting
amount of additional cooling due to adiabatic expansion.  Our cool
equilibrium temperature agrees well with the model atmospheres in
radiative and hydrostatic equilibrium by
\cite{anderson89c}
in which CO cooling reduces the temperature to 2640\,K.  It is also in
line with AA89, who, depending on the amount of prescribed mechanical
heating, find temperatures down to $\sim 3000$\,K, and the 2900\,K
already found by 
\citet{ayres81}.

The treatment of radiative transfer needs to be simplified in order to
keep the problem computationally tractable in the framework of a
multi-D radiation chemo-hydrodynamics simulation.  The use of only two
opacity bands is indeed a strong simplification, in particular in view
of the large number of spectral lines of different line strengths in
the CO fundamental vibration-rotational bands.
\citet{nordlund85} 
states that ``the cooling in the infrared CO lines is compensated for
by a heating in the ODFs in the blue and near UV'' if UV opacities
($\lambda < 400$\,nm) are treated as pure absorption, as in the
method applied here (based on MU85).  The alternative assumption of
pure scattering is questionable
\citep[cf. ][]{anderson89c}.
The fact that we still find a net cooling due to CO in the absence of
atmospheric dynamics can be interpreted such that our simplified LTE
treatment overestimates the cooling ability of the IR CO lines.
\citet{muchmore88} compare the scheme by MU85 with a more detailed
scheme with more frequency points and conclude that (i) the CO
overtone bands at 2.2\,$\mu$m play only a minor role compared to the
fundamental bands in the 5\,$\mu$m region and (ii) the CO cooling
might be overestimated by a factor of three with this simple scheme.

A similar conclusion can be drawn from the comparison of the
temperature stratifications from our extended cooling experiment with
the NLTE models by AA89 (see Sect.~\ref{sec:presheat}). It reveals
that the equilibrium temperature in the atmosphere is lower than found
by AA89 and that the amount of additional heating required to
compensate CO cooling is higher for our models.  Obviously, our
approach represents an upper limit for the net radiative cooling rate. 
So if even an over-efficient CO cooling mechanism cannot induce
significant effects on the atmospheric stratification, as shown in the
study presented here, then we can safely conclude that a more
realistic treatment with smaller net cooling will have even less
effect.

The inclusion of CO line opacity in our simulations indeed produced
additional radiative cooling at the front of fast-propagating shock
waves in the upper atmospheric layers due to a higher CO number
density compared to the instantaneous chemical equilibrium (ICE) case.
In contrast to the ICE case, the dissociation of CO proceeds on a
finite timescale so that the CO number density in the front of a
passing shock wave is reduced gradually and not instantaneously.
As a consequence, additional radiative cooling is present at these
locations.  The resulting changes in the thermal structure, however,
remain small.

It is commonly known that the dynamics tend to be too strong in 2D
compared to 3D. But even in our 3D simulations we do not find the
passage of shock waves to be less frequent, and the temperature
fluctuations are comparable in amplitude except for somewhat larger
differences in the middle photosphere
\citep{wedemeyerphd}.  
The 3D model by
\citet{wedemeyer05b} 
nevertheless yields a CO distribution that is very similar to the 2D
results presented here, confirming that the dynamics in the 2D model
are reasonably realistic.

We conclude that the self-amplifying cooling process of CO can in
principle operate in the solar atmosphere (as demonstrated in
Fig.~\ref{fig.coolexp} and by earlier works) and could lead to a
bifurcation between cool areas dominated by RE and warmer areas
dominated by mechanical heating. However, a relaxation to a cool state
of the solar atmosphere is prevented in the presence of pronounced
atmospheric dynamics as predicted by our models.  The crucial point is
that the CO radiative relaxation timescales are too long compared to
the frequent passage of shock waves.  Instead, the co-existence of hot
and cool regions is predominantly caused by mechanical heating due to
these propagating shock waves and adiabatic cooling of the resulting
post-shock regions 
\citep{wedemeyer04a}.  
Without any doubt, the present state-of-the-art simulations still
admit to some limitations, and the predictions of the velocity and
temperature fluctuations in the higher layers are thus affected by
uncertainties.  Detailed observational tests of atmospheric
properties, such as temporal and spatial temperature and velocity
variations and their centre-to-limb behaviour, are therefore highly
desired.

\begin{acknowledgements}
  We would like to thank T.~Ayres, J.~Bruls, O.~Steiner, and
  R.~J.~Rutten for discussion and helpful comments.  SWB is grateful
  to the Institut f\"ur Theoretische Physik und Astrophysik der
  Universit\"at Kiel for hospitality.  This work was supported by the
  {\em Deutsche Forschungs\-gemein\-schaft (DFG)}, project Ste~615/5.

\end{acknowledgements} 
\bibliographystyle{aa}

\end{document}